\documentclass[reprint,superscriptaddress,nofootinbib]{revtex4-2}
\usepackage{amsmath,amssymb}
\usepackage{mathptmx}
\usepackage{hyperref}
\usepackage{cleveref}
\usepackage{xurl}
\usepackage{graphicx}
\usepackage{xcolor}

\begin{document}

\title{Neural network predictions of plasma confinement loss in Wendelstein 7-X
       pellet-fueled discharges}

\author{K.~C.~Hammond}
\email{khammond@pppl.gov}
\affiliation{Princeton Plasma Physics Laboratory, Princeton, NJ, USA}

\author{J.-P.~B{\"a}hner}
\affiliation{Max Planck Institute for Plasma Physics, Greifswald, Germany}

\author{J.~Baldzuhn}
\affiliation{Max Planck Institute for Plasma Physics, Greifswald, Germany}

\author{S.~Bozhenkov}
\affiliation{Max Planck Institute for Plasma Physics, Greifswald, Germany}

\author{K.~J.~Brunner}
\affiliation{Max Planck Institute for Plasma Physics, Greifswald, Germany}

\author{A.~Dinklage}
\affiliation{Max Planck Institute for Plasma Physics, Greifswald, Germany}

\author{E. Edlund}
\affiliation{SUNY Cortland, Cortland, NY, USA}

\author{G.~Fuchert}
\affiliation{Max Planck Institute for Plasma Physics, Greifswald, Germany}

\author{M.~Huck}
\affiliation{Max Planck Institute for Plasma Physics, Greifswald, Germany}

\author{A.~I.~Mohammed}
\affiliation{Princeton Plasma Physics Laboratory, Princeton, NJ, USA}

\author{N.~Pablant}
\affiliation{Princeton Plasma Physics Laboratory, Princeton, NJ, USA}

\author{M.~Porkolab}
\affiliation{MIT Plasma Science and Fusion Center, Cambridge, MA, USA}

\author{G.~L.~Schmidt}
\affiliation{Princeton Plasma Physics Laboratory, Princeton, NJ, USA}

\author{J.~Smoniewski}
\affiliation{MIT Plasma Science and Fusion Center, Cambridge, MA, USA}

\author{T.~Stange}
\affiliation{Princeton Plasma Physics Laboratory, Princeton, NJ, USA}

\author{N.~Tamura}
\affiliation{Max Planck Institute for Plasma Physics, Greifswald, Germany}

\author{E.~Villalobos Granados}
\affiliation{Max Planck Institute for Plasma Physics, Greifswald, Germany}

\author{A.~von Stechow}
\affiliation{Max Planck Institute for Plasma Physics, Greifswald, Germany}

\author{G. Weir}
\affiliation{Max Planck Institute for Plasma Physics, Greifswald, Germany}

\author{the W7-X team}

\begin{abstract}

The energy confinement time is a key parameter of a magnetized fusion plasma,
helping to determine whether ignition can occur. 
Experiments in tokamaks and stellarators have shown that the 
confinement time can be improved via pellet injection.
The state of enhanced confinement brought about by a given pellet typically
deteriorates over time unless and until a subsequent pellet is injected.
In this work, we develop a data-driven model that predicts, at any moment, the
remaining time before a plasma in Wendelstein 7-X (W7-X) will lose its enhanced 
confinement state.
This ``remaining time'' metric effectively sets a deadline for when the next
pellet must be injected in order to steadily maintain a high confinement time.
We describe the development and training of the model and compare its 
predictions to observations from previous experiments. At least 90\% of the
model predictions are accurate to within 51~ms, which is below the 
typical W7-X energy confinement time as well as the minimum time separation
between subsequent pellet injections.
The model can be evaluated rapidly and could be suitable for use in a 
control system that optimizes the pellet injection rate in real time.

\end{abstract}

\maketitle

\section{Introduction}
\label{sec:intro}

It has long been understood 
\cite{lawson1957a,wurzel2022a} that the viability of nuclear fusion as an 
energy source depends in part on the energy confinement time $\tau_E$ of
the plasma in which the fusion takes place. In magnetic confinement concepts
such as the stellarator, the energy confinement time is limited by multiple 
energy loss mechanisms, including magnetohydrodynamic instabilities, 
neoclassical particle losses, and turbulent transport.  

Turbulent transport in particular has frequently been an impediment to attaining
high plasma performance in stellarators. To mitigate its impacts, two approaches
have been pursued. The first is to optimize the shaping of the magnetic field 
to stabilize the microinstabilities that lead to turbulent transport
\cite{mynick2010a,proll2016a,jorge2024a,kim2024a,goodman2024a,acton2024a}. 
The second is 
to optimize plasma control parameters such as heat and particle injection 
in order to achieve profiles of temperature and density that tend to suppress
turbulent fluctuations. 

Experiments in Wendelstein 7-X (W7-X) have demonstrated the promise of the 
second approach. On this device, the attainment of centrally-peaked density 
profiles with strong density gradients tends to 
result in lower turbulent fluctuations and reduced transport 
\cite{carralero2021a}. Such density gradients can be achieved via
wall conditioning methods such as boronization \cite{sereda2020a,ford2024a}
or powder injection \cite{lunsford2021a}, or via core fueling methods such
as neutral beam injection (NBI) \cite{ford2024a}
or hydrogen pellet injection \cite{baldzuhn2020a}. 

Pellet injection in particular has offered a promisng pathway to attaining
high plasma performance. Some of the early record triple 
products attained on W7-X were enabled by a combination of electron cyclotron
resonance heating (ECRH) and pellet fueling 
\cite{pedersen2019a,bozhenkov2020a}. W7-X is currently equipped
with a continuous pellet fueling system with the capability to inject
pellets for arbitrary amounts of time 
\cite{meitner2020a,meitner2022a,meitner2023a}, making pellet injection the only
core fueling option that can be employed continuously for long 
pulses ($>$30 s) on W7-X. So far, the pellet injector has maintained elevated
performance parameters for durations of at least 40 seconds \cite{grulke2026a}.

To date, pellets in W7-X have only been injected at pre-programmed times
during plasma discharges. However, those injection times have not necessarily
been optimal for maintaining high performance. 
It is likely that performance could be substantially improved if 
the pellet injections could be regulated in real time by a feedback control
system. Such a system would choose the optimal times for injected pellets
according to the observed plasma parameters, making adjustments based on
evolving plasma conditions that would be difficult to predict in advance.
Feedback control capabilities for pellets
are important to develop not just for W7-X but for stellarators
as a whole, as many current designs for stellarator power plants 
envision pellet injection as a key source of fueling and profile control
\cite{lion2025a,guttenfelder2025a,hegna2025a,swanson2026preprint}.

In this paper, we describe an early step toward profile control by developing a
simple data-driven neural network model whose predictions can be used in 
real time to inform
a control system on a choice of pellet injection rate. Specifically, the model
uses real-time available data on the plasma temperature and density profiles
to predict how much longer an ECRH-heated plasma can maintain a state of high 
performance without a subsequent pellet injection. This quantity, defined as
the ``remaining time'' $t_r$, effectively sets a minimum required pellet 
injection rate for maintaining consistently high performance parameters.

The paper is organized as follows. Sec.~\ref{sec:pellet_injection} 
describes the basic characteristics of pellet injection as observed on W7-X that
were used to inform the model definition. Sec.~\ref{sec:model_dev} describes
the development and training of the model. In Sec.~\ref{sec:feedback_potential},
we input signals from previous discharges with pellet injection to the model
to demonstrate how the incorporation of the model's predictions into a 
feedback system could have improved the performance of those discharges.
Sec.~\ref{sec:discussion} discusses possible experimental use cases as well
as opportunities for future improvements to the model.

\section{Characteristics of pellet injection in Wendelstein 7-X}
\label{sec:pellet_injection}

\begin{figure}
    \includegraphics[width=0.48\textwidth]{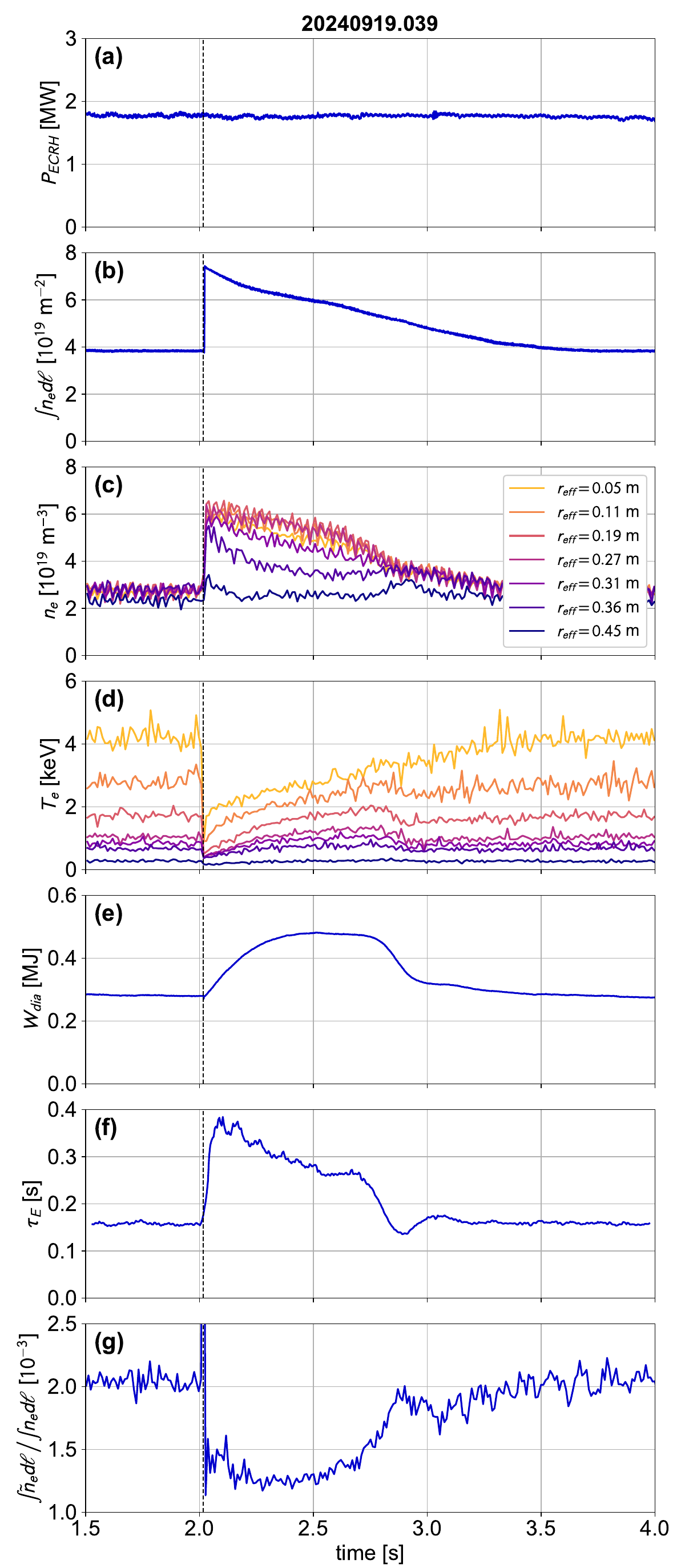}
    \caption{Time traces of key experimental parameters during the injection of
             a pellet in a low-density plasma:
             (a) total ECRH heating power,
             (b) line-integrated electron density,
             (c) electron density at selected effective minor radii 
                 $r_\text{eff}$,
             (d) electron temperature at selected $r_\text{eff}$,
             (e) plasma stored energy,
             (f) energy confinement time, 
             (g) normalized line-integrated density fluctuation amplitude 
                 over the frequency band of 50-1000 kHz.
             The vertical dashed line indicates the approximate time at which
             the pellet arrives at the plasma edge.
            }
    \label{fig:pellet_example_traces}
\end{figure}

The predictive model developed in this paper relies partly on the tendency
for the plasma parameters in Wendelstein 7-X to exhibit distinctive temporal
evolution patterns in response to the injection of pellets. These patterns
are illustrated in Fig.~\ref{fig:pellet_example_traces} for the injection
of a single pellet. In this example, the pellet was injected into a
relatively low-density plasma heated with a relatively low and constant level of
ECRH power, $P_{ECRH}$ (Fig.~\ref{fig:pellet_example_traces}a). No other heating
sources were employed in this discharge. More broadly, in this work, only 
plasmas heated with pure ECRH are considered.

The pellet has a strong and immediate impact on the plasma density, as 
illustrated in Fig.~\ref{fig:pellet_example_traces}b. This shows the time
behavior of the line-integrated electron density as measured by the dispersion
interferometer \cite{brunner2018a}. The vertical dashed line indicates the
approximate time at which the pellet first strikes the plasma at approximately
$t$ = 2 s. Immediately, the line-integrated density nearly doubles as the 
pellet deposits its mass. Then, more gradually, the density
decays over the course of 1-2 seconds back to its pre-pellet level.

The pellet impacts not just the overall plasma density but also how the density
is spatially distributed. This is apparent from 
Fig.~\ref{fig:pellet_example_traces}c, which shows measurements of the electron
density $n_e$ by the Thomson scattering diagnostic 
\cite{pasch2016a,bozhenkov2017a}
at selected points of differing effective minor radius $r_\text{eff}$.
The vertical spacing between the time traces $n_e$ at different spatial points 
in Fig.~\ref{fig:pellet_example_traces}c indicates
how the density profile, defined as $n_e(r_\text{eff})$, evolves over time.
A large vertical separation between $n_e$ traces at nearby spatial locations
indicates a large radial gradient in $n_e$.

Prior to the pellet injection ($t < 2$ s), $n_e$ is approximately the same at
all minor radii, indicating that the density is uniform throughout the plasma
volume; i.e. the profile is \textit{flat}. Once the pellet is injected, $n_e$ 
at most of the spatial points rises quickly with 
$n_e$ at $r_\text{eff} < 0.4$ m reaching
double their initial values on a sub-millisecond time scale. This implies that
the pellet mass has been deposited deep in the core of the plasma and is 
regularly observed in discharges heated with pure ECRH. This is a 
surprising result that is not expected from modeling, which tends to predict
shallower deposition 
\cite{panadero2018a,baldzuhn2019a}, and is currently under active investigation
\cite{damm2026a}. 

Shortly after injection of the pellet ($t \gtrsim 2$ s), a spatial gradient
appears in the outer radii ($r_\text{eff} \gtrsim 0.3$ m).
Due to the presence of the spatial gradient, the profile at this time is 
considered to be \textit{peaked}. Then, over the next 1-2 s, the densities 
slowly decrease, at different rates at different locations. Finally, shortly 
before 3 s, the density traces converge back together, and the profile is back 
to being flat.

The pellet impacts the temperature profile as well as the density profile.
Fig.~\ref{fig:pellet_example_traces}d shows time traces of electron temperature
$T_e$ measured by the Thomson scattering diagnostic at the same points
for which density is shown in Fig.~\ref{fig:pellet_example_traces}c. The 
immediate impact of the pellet injection is to decrease $T_e$, most dramatically
near the plasma core ($r_\text{eff} \lesssim 0.25$ m). The temperatures 
gradually recover to their pre-pellet values on a time scale similar to that 
of the density decay. Notably, the $T_e$ profiles always exhibit spacial 
gradients near the plasma core, although they are reduced in the aftermath of 
the pellet.

The pellet clearly has a beneficial impact on energy confinement, as can be
seen from Fig.~\ref{fig:pellet_example_traces}e-f.
Fig.~\ref{fig:pellet_example_traces}e shows the plasma stored energy 
$W_\text{dia}$ obtained from the diamagnetic loop \cite{rahbarnia2018a}.
Prior to the pellet injection, $W_\text{dia}$ maintains a constant value.
After the pellet is injected, $W_\text{dia}$ gradually increases until 
reaching a maximum value around $t = 2.5$ s. The fact that the stored energy
increases while the input heating power remains constant indicates that
the overall energy confinement capability increases as well. This improved
confinement does not last forever, though; eventually, $W_\text{dia}$ begins
to decay to its pre-pellet level shortly before $t = 3$ s.

The confinement quality can be quantified more precisely by the energy 
confinement time $\tau_E$, shown in Fig.~\ref{fig:pellet_example_traces}f.
In this work, $\tau_E$ is estimated from $P_{ECRH}$ and $W_\text{dia}$ as
follows:

\begin{equation}
\label{eqn:tau_E}
    \tau_E = \frac{W_\text{dia}}{P_{ECRH}-dW_\text{dia}/dt}
\end{equation}

\noindent Here, the time derivative of $W_\text{dia}$ is computed through finite 
differencing, and for this purpose $W_\text{dia}$ is smoothed through boxcar 
averaging with a time window of 0.1 ms to mitigate the impact of noise on the
differentiation. Since $W_\text{dia}$ and $P_\text{ECRH}$ are obtained from
different diagnostics on different timebases, the respective signals are
interpolated to compute $\tau_E$ at defined time points. We also note that
the heating power term in the equation is by definition the heating
power absorbed by the plasma. By contrast, the signal used here for $P_{ECRH}$ 
is the \textit{launched} ECRH power, not all of which is absorbed. However, ECRH 
absorption is typically
greater than 90\% in W7-X, even at higher densities when second-harmonic
O-mode polarization is employed \cite{wolf2019b,laqua2021a}; hence $P_{ECRH}$ 
is a sufficiently accurate estimate of absorbed power for the purposes of the
analysis in this paper.

As shown in Fig.~\ref{fig:pellet_example_traces}f, $\tau_E$ quickly rises upon
pellet injection, reaching a value that is nearly double its pre-pellet
value. The greatly elevated confinement time corresponds to the plasma's 
observed ability to accumulate stored energy at a constant heating
power (Fig.~\ref{fig:pellet_example_traces}e). For about one second after
the pellet is injected, $\tau_E$ decays gradually. Then, around $t = 2.8$~s,
it drops precipitously to its pre-pellet level, roughly at the same time at
which the $n_e$ profile flattens (Fig.~\ref{fig:pellet_example_traces}c).

Such a temporary phase of enhanced energy confinement is regularly observed
in the wake of pellet injection in W7-X \cite{bozhenkov2020a,baldzuhn2020a}.
These phases are associated with reduced density fluctuation levels
\cite{estrada2021a,carralero2021a}, indicating that the increased confinement
can be explained by a reduction of losses due to turbulent transport.
The reduced turbulence is believed to be caused by the presence of increased 
density gradients brought about by the pellet's mass deposition, which tend
to stabilize the microinstabilities that drive the ion temperature gradient
mode in W7-X \cite{alcuson2020a,thienpondt2025a}.
Also during these phases, ion temperatures tend to equilibrate with the
electron temperatures \cite{bozhenkov2020a,baldzuhn2020a} and reach levels 
higher than what is attainable
in ECRH-heated discharges without pellet fueling \cite{beurskens2021a}.

The interpretation of improved confinement through reduced turbulence is 
supported by measurements of density fluctuations, as shown in 
Fig.~\ref{fig:pellet_example_traces}g. Plotted here is the amplitude of the
line-integrated density fluctuations integrated over the frequency band 
50-1000 kHz as measured by the Phase Contrast Imaging (PCI) diagnostic 
\cite{edlund2018a,huang2021a}, normalized to the
line-integrated electron density. The time trends of this quantity are 
qualitatively opposite to those of $\tau_E$, featuring a quick drop after 
the pellet is injected and a return to the pre-pellet level by the time
the density profiles have become flat again. The reduced normalized fluctuation
level observed during the phase of enhanced confinement is consistent with
reduced turbulent transport and has been found to be correlated with density
profile peaking \cite{grulke2026a,vonstechow2025apsdpp}.

\begin{figure*}
    \includegraphics[width=\textwidth]{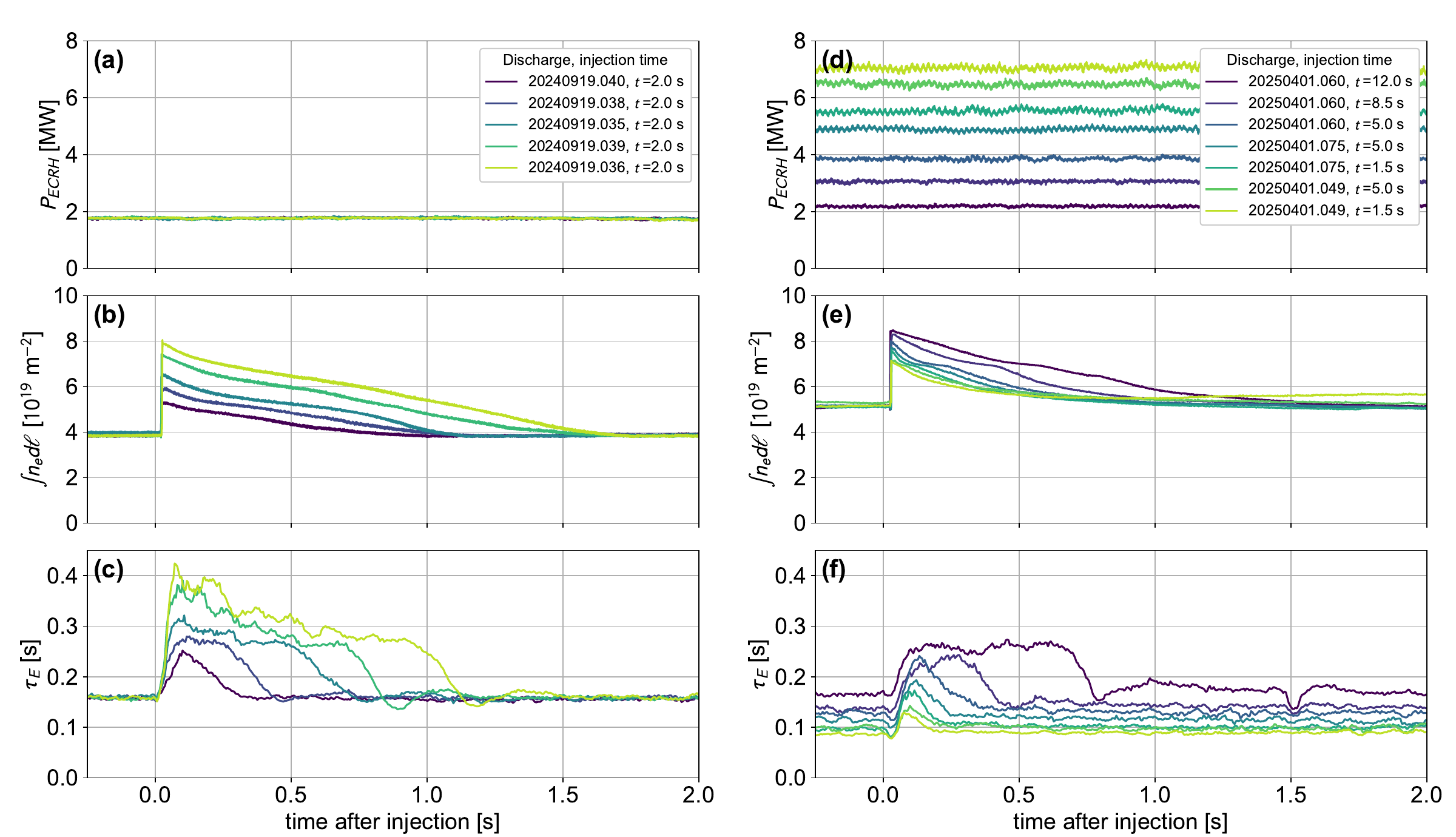}
    \caption{Times trends of key plasma parameters after sets of pellet 
             injections in which the pellet mass was varied ((a)-(c)) and 
             in which the ECRH power was varied ((d)-(f)). 
             (a) and (d): ECRH power, 
             (b) and (e): line-integrated electron density, and
             (c) and (f): energy confinement time.
             Note: the discharges shown in (a)-(c) were performed in the
             high-mirror magnetic configuration, whereas the discharges in
             (d)-(f) were performed in the low-iota magnetic configuration.}
    \label{fig:enhanced_confinement_duration}
\end{figure*}

The duration of the period of enhanced confinement following a pellet injection
varies considerably according to the pellet parameters and the underlying 
plasma conditions. Two frequently-observed qualitative trends are shown in
Fig.~\ref{fig:enhanced_confinement_duration}. The first column
(Fig.~\ref{fig:enhanced_confinement_duration}a-c) shows traces of key
plasma parameters in the wake of injections of pellets of different masses.
While calibrated measurements of the pellet mass are not currently available,
the relative masses of the pellets are qualitatively indicated by the amount
by which the line-integrated density rises upon injection, with the largest
pellets producing the greatest rise.
Other things being equal, larger pellets generally result in longer periods
of enhanced confinement. For the largest pellet in the set, the phase of 
enhanced confinement lasted for more than 1 s, whereas for the smallest pellet,
the phase lasted for only about 250 ms. In addition, the maximum increase in
$\tau_E$ was greater for larger pellets.

Another parameter that was consequential for the enhanced confinement phase was
the level of ECRH power employed.
Fig.~\ref{fig:enhanced_confinement_duration}d-f shows traces of plasma
parameters in the wake of injections of pellets of
similar masses but with different levels of (constant) ECRH power. As heating
power increases, both the duration of the phase of enhanced confinement 
and the maximum increase in $\tau_E$ tend to 
decrease with heating power. The baseline confinement time also decreases
with heating power, as is expected from the ISS04 scaling law 
\cite{yamada2005a}.

The data shown so far have been from injections of single, isolated pellets.
In these cases, the pellet induces a temporary phase of enhanced performance
that eventually relaxes back to the lower-performance state that the plasma
had been in before the pellet was injected. However, if multiple pellets are
injected rapidly in sequence, it is possible to sustain the enhanced performance
phase for a longer period of time. The key requirement for sustaining the
enhanced performance from one pellet to the next is that the subsequent pellet
should be injected before the relaxation occurs, i.e. before $\tau_E$ drops
to its pre-pellet level.

\begin{figure}
    \includegraphics[width=0.49\textwidth]{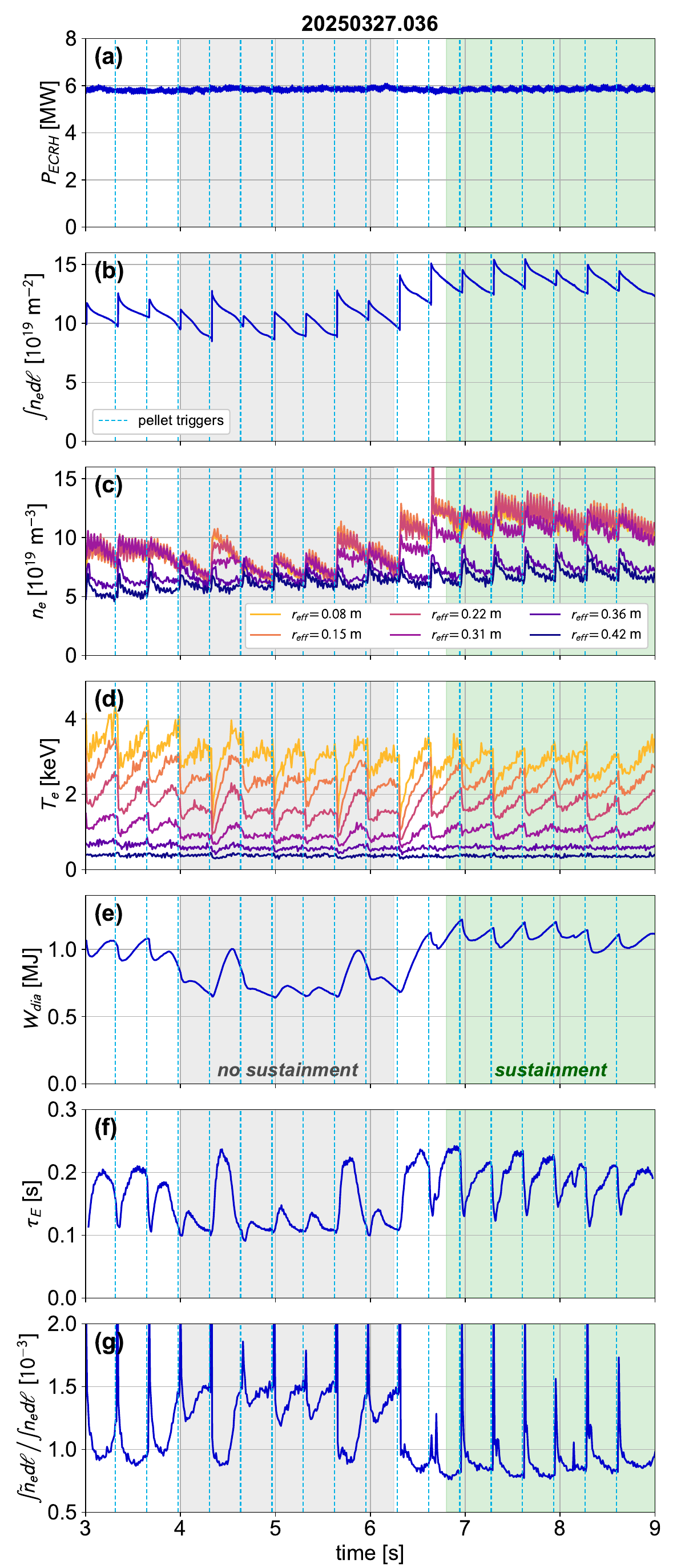}
    \caption{Time traces of key plasma parameters during a series of pellets
             during which enhanced confinement is sustained at some times but
             not others.
             (a) ECRH power,
             (b) line-integrated electron density,
             (c) electron density at selected $r_\text{eff}$,
             (d) electrom temperature at selected $r_\text{eff}$,
             (e) diamagnetic energy,
             (f) confinement time, and
             (g) normalized fluctuation amplitude for 50-1000 kHz.
             Vertical dashed lines indicate the trigger times for launching 
             pellets. The gray box indicates a period of time when enhanced
             confinement is \text{not} sustained; the green box indicates
             a period when enhanced confinement is successfully sustained.}
    \label{fig:series_varying_sustainment}
\end{figure}

Fig.~\ref{fig:series_varying_sustainment} shows a discharge in which a pellet
series successfully maintains enhanced confinement at some times but not at
others. During the time interval shown, the plasma is heated by ECRH at a 
steady level of 6 MW (Fig.~\ref{fig:series_varying_sustainment}a). 
Pellets are injected at a constant rate of approximately 3 Hz, with the 
vertical dashed lines indicating the trigger times for each pellet.

The shaded regions of Fig.~\ref{fig:series_varying_sustainment} highlight
time intervals in which the pellet series has different levels of success
at sustaining enhanced confinement. During the time interval shaded in gray
(from 4 to 6.25 s), enhanced confinement is \textit{not} sustained, whereas
during the interval shaded in green (from 6.8 to 9 s), enhanced confinement
is successfully sustained. The difference in plasma performance can be seen
clearly in the time trace of diamagnetic energy $W_\text{dia}$
(Fig.~\ref{fig:series_varying_sustainment}e). During the interval 
\textit{without} sustainment (gray), $W_\text{dia}$ assumes lower values, 
mostly around 0.6 MJ, whereas during the interval \textit{with} sustainment 
(green), $W_\text{dia}$ is mostly above 1 MJ. In addition, the normalized 
fluctuation amplitude measured by the 
PCI diagnostic \cite{edlund2018a,huang2021a,vonstechow2025apsdpp} 
(Fig.~\ref{fig:series_varying_sustainment}g) tends to be 
greater during the no-sustainment interval than during the sustainment 
interval, indicating that turbulence is more consistently suppressed during the 
latter. 

Close inspection of the evolution of the energy confinement time $\tau_E$ 
(Fig.~\ref{fig:series_varying_sustainment}f) shows a key difference between the 
two intervals. During the interval of no sustainment (gray), each pellet
injection (indicated by a vertical dashed line) prompts an immediate rise in 
$\tau_E$. However, $\tau_E$ falls back roughly to its initial level before
the subsequent pellet is injected. By contrast, during the interval of
sustainment (green), pellet injections occur while $\tau_E$ is still
elevated (i.e. it has not spontaneously fallen to a pre-pellet level).
Interestingly, the injection of pellets during this phase appears to prompt
an immediate and rapid drop in $\tau_E$ along with a spike in the fluctuation
amplitude (Fig.~\ref{fig:series_varying_sustainment}g). However, during this
interval, the plasma quickly recovers from the pellet-induced drops in
$\tau_E$, resulting in a higher time-averaged $\tau_E$ than during the 
no-sustainment interval and a persistantly higher $W_\text{dia}$
(Fig.~\ref{fig:series_varying_sustainment}e). Overall, this supports the
interpretation that enhanced plasma confinement may be sustained by a series
of pellets as long as each pellet is injected soon enough, before the plasma
relaxes to its lower-confimenent pre-pellet state.

To understand why sustainment would vary during a discharge at which pellets
were injected at a constant rate, it is important to note 
that while the rate was fixed, the pellet mass varied from one
pellet to the next. Injection of larger pellets, as observed in 
Fig.~\ref{fig:enhanced_confinement_duration}, tends to result in longer-lived 
periods of enhanced confinement. 
We posit that this effect helped to bring about the transition from no 
sustainment to sustainment in the discharge depicted in 
Fig.~\ref{fig:series_varying_sustainment}.
In the transition between the no-sustainment interval and 
the sustainment interval (from 6.25 to 6.8 s), two especially large pellets
were injected, as indicated by their larger rises in line-integrated density
(Fig.~\ref{fig:series_varying_sustainment}b) and core density (the yellow and
orange traces in Fig.~\ref{fig:series_varying_sustainment}c). These larger
pellets appear to have induced phases of enhanced confinement that were long
enough that the subsequent pellets arrived before the relaxation of the 
plasma.

While the pellets leading into the sustainment phase were larger than most of 
the other pellets in the discharge, it is notably \textit{not} the case that
all pellets during the interval of sustainment (green) were equally large,
comparable to the pellets injected during the no-sustainment interval (gray).
Yet during the sustainment interval, the plasma does not relax back to its
low-confinement state between pellets as it does during the no-sustainment
interval. However, there is another significant difference between these
two intervals, which is that the core electron density 
(Fig.~\ref{fig:series_varying_sustainment}c) is substantially greater during 
the sustainment interval than during the no-sustainment interval. Furthermore,
the degree of density peaking, indicated by the separation between the core and
edge densities, is consistently larger than in the no-sustainment interval.
Given the association between density peaking and reduced transport in W7-X
\cite{vonstechow2025apsdpp},
it may be the case that greater levels of density profile peaking also lead 
to longer periods of enhanced confinement after pellet injection. 

While the foregoing discussion offers some speculative interpretation of the
varying plasma behaviors observed in the discharge in 
Fig.~\ref{fig:series_varying_sustainment}, it should be noted that these
phenomena are not fully understood and currently the subject of active study.
However, even in absence of a full physical understanding, enough data have 
been collected to date for the development of phenomenological models that 
can make predictions based on observed patterns. We describe the development
of one such model in the next section.


\section{Model development}
\label{sec:model_dev}

Given the extensive variability in the amount of time it takes for the plasma
to lose enhanced confinement, it would be useful for a plasma control system
to be able to predict this according to diagnostic signals. Such a prediction
would help to inform the control system at any given time how soon it would
need to inject a pellet in order to keep the plasma in the state of enhanced
confinement. This would enable the system to maintain high performance more
reliably for long plasma discharges, particularly in the face of unplanned
changes in pellet parameters, heating power, and wall conditions.

To this end, we have formulated a model that takes real-time plasma diagnostic
data as input and outputs a prediction of the \textit{remaining time}, $t_r$, 
during which the plasma is expected to stay in a state of enhanced confinement
before relaxing to a lower-confinement state in absence of a subsequent 
intervention such
as a pellet injection. For example, if at a given moment the prediction 
for $t_r$ is 100~ms, the implications are that (1) the plasma is presently in a 
state of enhanced confinement and (2) it will relax back to a state of 
(pre-pellet) lower confinement in 100 ms without a subsequent pellet assuming 
a constant level of heating power. This effectively provides a deadline for
injecting a subsequent pellet if the objective is to keep the plasma in the
state of enhanced confinement.

The model is based on an ansatz that presumes that, for a fixed magnetic 
configuration and with constant ECRH heating power, $t_r$ can be determined
from the profiles of electron temperature and density as well as
the heating power:

\begin{equation}
    \label{eqn:t_r_ansatz}
    t_r = t_r(T_e(r_\text{eff}), n_e(r_\text{eff}), P_{ECRH})
\end{equation}

\noindent This seems reasonable based on the general observation that energy
transport in W7-X plasmas is often dominated by turbulence
\cite{beidler2021a,beurskens2021a}, which in turn depends strongly on the 
radial gradients of 
temperature and density that are encoded in the profiles. The inclusion of 
the heating power as an explicit input is informed by observations of its 
influence on the durations of pellet-induced phases of enhanced confinement
as seen in Fig.~\ref{fig:enhanced_confinement_duration}d-f.

The inputs to the model are thus the values of electron temperature 
$T_e$ and electron density $n_e$ at a set of points with different
effective minor radii $r_\text{eff}$, as well as the total ECRH
heating power $P_{ECRH}$.
$T_e$ and $n_e$ are supplied by the Thomson scattering
diagnostic by way of an evaluation system that has recently been commissioned
to process the data in real time \cite{mohammed2026a}. This system produces
updated profiles of $T_e$ and $n_e$ roughly every 11~ms.
The $n_e$ values may be scaled with real-time-calculated values of 
line-integrated density to help mitigate the impact of systematic errors in
the Thomson density calibration \cite{fuchert2022a,nelde2023a}.
The values of $P_{ECRH}$ are provided by radio-frequency diodes in the ECRH 
transmission lines and are updated much more frequently (25 kHz).

\subsection{Curation of a training dataset}
\label{sec:training}

\begin{figure}
    \includegraphics[width=0.49\textwidth]{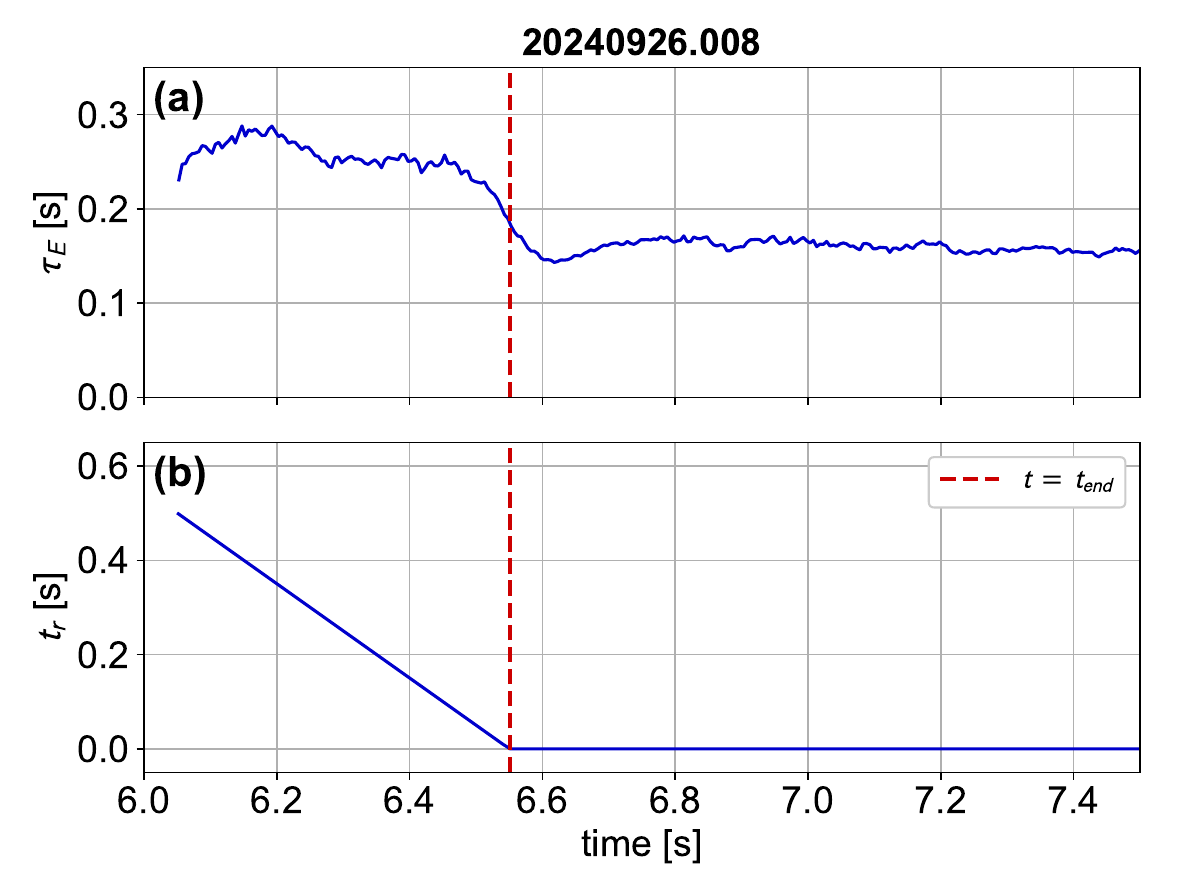}
    \caption{Example time traces illustrating how $t_r$ is defined for 
             previously-observed plasma data, recorded over an interval 
             beginning just after the injection of a pellet:
             (a) energy confinement time and
             (b) remainingn time. 
             The vertical dashed line indicates $t_\text{end}$ as defined
             in the text.}
    \label{fig:tr_definition}
\end{figure}

The model was trained using prior experimental
data in which the model inputs were known and the true value of $t_r$ 
associated with those inputs could be determined.
The key signal used to determine $t_r$ from previous discharges was the 
energy confinement time $\tau_E$.
Fig.~\ref{fig:tr_definition}
illustrates how $t_r$ is defined for an example set of experimental data.
Fig.~\ref{fig:tr_definition}a shows a time trace of 
$\tau_E$ for an interval of a plasma discharge that began in a state of
enhanced confinement and relaxed over time to the pre-pellet state. In this
case, the beginning of the interval is shortly after the injection of a pellet,
which induced the enhanced confinement. The end of the period of enhanced
confinement can be identified by the curve of $\tau_E$ falling and reaching
a steady value.

The value of $t_r$ is then defined according to the observed evolution of 
$\tau_E$. Fig.~\ref{fig:tr_definition}b shows the values $t_r$ for each 
time point in the interval corresponding to the $\tau_E$ curve in 
Fig.~\ref{fig:tr_definition}a. First, a single time point $t_\text{end}$ 
corresponding to the end of the phase of enhanced confinement is 
identified. For simplicity and consistency,
we have defined this as the point at which $\tau_E$ falls to 
$\tau_{E,\text{max}} - 0.75(\tau_{E,\text{max}} - \tau_{E,\text{final}})$,
where $\tau_{E,\text{max}}$ is the maximum value of $\tau_E$ observed during the
interval and $\tau_{E,\text{final}}$ is the final value of $\tau_E$ in the
interval. Then, $t_r$ for each time point in the interval is defined as follows:

\begin{equation}
    t_r(t) = 
        \begin{cases}
            t_\text{end} - t, & t \leq t_\text{end} \\
            0,                & t > t_\text{end}
        \end{cases}
    \label{eqn:t_r}
\end{equation}

\noindent In other words, for all time points prior to the end of the 
enhanced confinement phase, $t_r$ is the amount of time before $t_\text{end}$,
whereas after the end of the phase, when the plasma is back in the pre-pellet
state of lower confinement quality, $t_r$ is defined as 0.

For past data to be useful for model training purposes, it had to come from
time intervals during which the true value of $t_r$ was known and well-defined.
This entailed certain criteria when assembling the training dataset.
First, as the model is intended for discharges heated exclusively with ECRH,
discharges using other heating sources such as neutral beam injection (NBI)
were excluded. Cases in which short NBI pulses (blips) were used for diagnostic
purposes were also excluded, as this would have perturbed the $\tau_E$ signal
and complicated the determination of $t_\text{end}$. Second, the plasma had
to be under conditions of steady actuation, with no large or sudden 
perturbations. To this end, while time intervals used for training purposes
typically included the aftermaths of pellet injections, they
excluded the $< 1$~ms window during which the ablation and deposition processes
took place. Other large perturbations that may or may not have been 
intentional, such as massive impurity injections, were also excluded.
Furthermore,
the ECRH power level needed to remain the same throughout the interval.
These exclusions were made because the model is intended to predict the
spontaneous evolution of the plasma properties via internal transport 
processes, and not externally controlled actuation events.
Finally, the interval had to be long enough for the plasma
to lose enhanced confinement and return to the pre-pellet phase; otherwise,
$t_\text{end}$ would not be known and $t_r$ could not be determined for any
point in the interval.

Another important consideration when selecting data for model training is the
magnetic configuration employed when the data were collected. The magnetic
configuration in W7-X is determined by the currents employed in the different
types of electromagnetic coils that generate the confining magnetic field. 
For a fixed set of scattering volumes used for evaluating
$T_e$ and $n_e$, it is likely that the same model inputs would exhibit different
$t_r$ in different magnetic configurations. One reason for this is that 
different configurations have different flux surface geometry; hence, the
scattering volumes where $T_e$ and $n_e$ are measured would have different
values of $r_\text{eff}$. In addition, different configurations exhibit
substantially different transport properties 
\cite{geiger2015a,dinklage2018a,stroteich2022a}. 

For the present work, rather than attempting to train a single model to 
distinguish between different magnetic configurations, we have opted to 
train separate models for different configurations. Accordingly, to train
any given model, we would collect a set of training data from plasma discharges
performed in one specific configuration. These models would then only
be applicable to make predictions for the respective configurations on which
they were trained. This is a workable approach in practice for W7-X, since
the majority of plasma experiments are performed in one of a small number
of configurations. 
For this paper, we have trained two models: one for the
``standard'' magnetic configuration, and one for the ``high-iota'' 
configuration. 


The training datasets for each model were selected from plasma discharges from
the OP2.2 and OP2.3 experiment campaigns. For the standard configuration 
model, the training data were drawn from 145 post-pellet time intervals from
39 discharges. For the high-iota configuration model, the data were drawn from
44 intervals from 10 discharges. The ECRH power levels applied during the 
intervals ranged from 1.2 to 6.5 MW for the standard configuration and from
2.7 to 7.5 MW for the high-iota configuration. The time intervals were selected 
according to the criteria discussed above, with each beginning in a state of 
enhanced confinement and ending after the plasma had relaxed to a steady state 
without enhanced confinement. 

One additional distinction between the two 
datasets is that the training data from the standard configuration was taken
from plasmas heated by ECRH launched in the second-harmonic extraordinary mode
(X2), whereas the data for the high-iota configuration was taken with plasmas
heated by ECRH launched in the second-harmonic ordinary mode (O2). The two
modes tend to exhibit different absorption profiles, which may impact 
transport in different ways. The O2 mode is more relevant for high-performance
discharges because it can be injected at higher plasma densities. However,
we opted to use data with X2 absorption for the data from the standard 
configuration due to a lack of useable data from plasmas heated in O2 mode
to date in that configuration.

It should be noted that the training sets did not explicitly control
for all possible factors that might effect profile evolution. For example,
many discharges employed gas puffing to fuel the plasma from the edge. 
The puffing rate in most cases was regulated by a control system designed
to maintain a consistent line-integrated density, and was not constant across
intervals. Another potentially important factor is impurity seeding, which 
is often employed to induce divertor detachment and reduce overall heat
loads to the divertor \cite{effenberg2019a}. Most of the training intervals
did not involve impurity seeding; however, this is likely to be present
in future scenarios where the model will be employed. 

Data from these time intervals were packaged into sample sets. Each sample
consisted of (1) $T_e$ and $n_e$ profiles
from the Thomson scattering diagnostic with $n_e$ scaled by the line-integrated
density from the interferometer, (2) the total ECRH heating power at 
the time that the $T_e$ and $n_e$ profiles were acquired, and (3) the
actual value of $t_r$ determined for the same time from the time trace of 
$\tau_E$. The overall sample set for the standard configuration containe
8,141 samples, whereas the sample set for the high-iota configuration contained
2,577 samples. 

\subsection{Model structure}
\label{sec:model_structure}

\begin{figure}
    \includegraphics[width=0.49\textwidth]{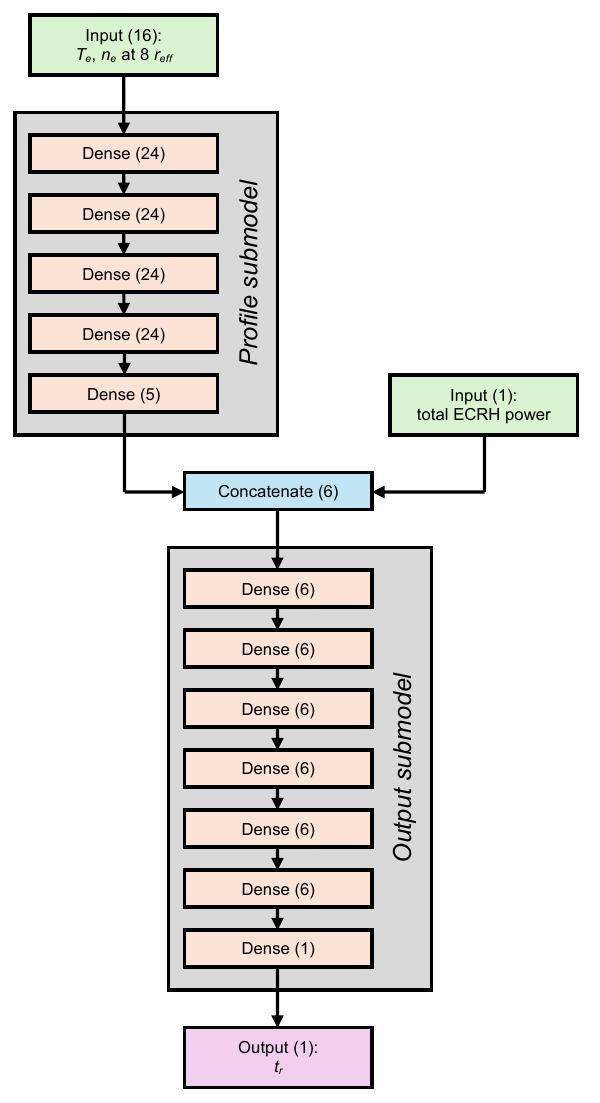}
    \caption{Schematic of the neural network model used to calculate $t_r$
             based on $T_e$ and $n_e$ profiles as well as the ECRH power.
             Individual layers in the model are depicted as colored boxes,
             with the numbers in parentheses specifying the dimensionalies
             of the respective layers. The gray boxes, which group layers 
             together, denote the submodels described in the text.}
    \label{fig:nn}
\end{figure}

The model is implemented as an artificial neural network with a structure as 
depicted in Fig.~\ref{fig:nn}. The model consists of two submodels: a
``profile'' submodel and an ``output'' submodel. The profile submodel
takes an electron temperature and density profile from the Thomson diagnostic
as input and outputs five values. The purpose of this submodel is thus to
distill the profiles down to a small set of feature parameters that are most
relevant for the purpose of determining $t_r$. These features are then
combined with the total ECRH heating power and
fed into the output submodel. The output
submodel calculates the final output of the full model; namely, the 
estimate of the remaining time $t_r$.

The model was constructed and 
trained using the Keras framework \cite{keras} with the TensorFlow backend
\cite{tensorflow}.
The data input to the profile submodel consist of 16 values:
the electron temperatures $T_e$ and 
densities $n_e$ obtained from eight scattering volumes of the 
Thomson scattering diagnostic with distinct effective minor radii 
$r_\text{eff}$. 
These values are input to a series of five dense, fully-connected layers. The
first four of these layers have 24 nodes, whereas the final layer has
five nodes corresponding to the output size. The output values are 
concatenated with the ECRH power value and fed into the output submodel.
The output submodel consists
of a sequence of seven dense, fully-connected layers, the first six of which
have six nodes and the last of which has one node. All layers used in the 
model employ the ``leaky RelU'' activation function.
 
It was not obvious \textit{a priori} what the best choices would be for various
structural aspects of the model. For example, it was unknown how many 
feature parameters are necessary to extract from the profiles via the profile
submodel to 
enable accurate predictions of $t_r$. To inform the selection of this and
other hyperparameters, we constructed and evaluated a set of candidate
model structures with different hyperparameter settings.
The hyperparameters under consideration are summarized in 
Table~\ref{tab:hparam_values}. For the profile submodel, hyperparameters 
included the number of layers, the number of nodes per layer in all but the 
last layer, and the number of nodes in the final layer. For the output
submodel, hyperparameters included the number of layers and the number of
nodes per layer in all but the last layer. Each hyperparameter was given
a range of five possible values, yielding a search space of $5^5 = 3,125$
possible model structures. We evaluated a total of 320 candidate model 
structures with hyperparameters selected
randomly from their corresponding value ranges, constituting a little over 
10\% of the overall hyperparameter search space.

\begin{table}
    \begin{tabular}{l l}
        \hline
        \textbf{Parameter} & \textbf{Values} \\
        \hline
        \textit{Profile submodel:} & \\
        ~~~~Number of layers & 2, 3, 4, 5, 6 \\
        ~~~~Nodes per layer (except last layer) & 16, 24, 32, 40, 48 \\
        ~~~~Nodes in last layer & 1, 2, 3, 4, 5 \\
        \textit{Output submodel:} & \\
        ~~~~Number of layers & 3, 4, 5, 6, 7 \\
        ~~~~Nodes per layer (except last layer) & 2, 3, 4, 5, 6 \\
        \hline
    \end{tabular}
    \caption{Candidate values for the structural aspects of the neural 
             network model considered for the hyperparameter search.}
    \label{tab:hparam_values}
\end{table}

For each candidate structure, models were fit to a fixed set of training 
data. The training set consisted
of a random selection of 60\% of the samples from the high-iota sample set.
The fits were performed with the Adam optimization algorithm \cite{kingma2015a},
which adjusts the model weights to minimize a loss function equal to the mean 
of the squares of the
errors (MSE) in the predictions of $t_r$ relative to the true values.
Because the fitting algorithm is non-deterministic, multiple models with
the same structure were fit independently 
to the same data to determine statistics for achievable fit
quality. Hererafter, a \textit{fit} of a candidate model structure will refer 
to a set of weights determined for that structure by the optimizer. 

The quality of each fit was quantified by the \textit{validation MSE}
$E_\text{val}$, or the MSE in predictions of
$t_r$ relative to a \textit{validation set} consisting of half of the high-iota
samples that were not used in the training set. For each candidate model,
the overall suitability was quantified by a \textit{score} consisting of
the average $E_\text{val}$ obtained from all the fits performed for
that model structure. The search was conducted using 
KerasTuner \cite{keras_tuner};
more details on the computation of the score can be found in the associated
documentation. 

During the search, one fit was retained for each candidate model structure. 
This fit was the one that had the lowest $E_\text{val}$ among all fits 
performed for the candidate structure. 
For this retained fit, an additional metric,
the \textit{test MSE}, $E_\text{test}$, was computed. 
$E_\text{test}$ was the MSE for predictions by the fit of $t_r$ for a 
\textit{test set} of samples,
consisting of all samples from the high-iota sample set that were not part 
of the training set or the validation set. 
The test MSE is useful as an additional metric for evaluating candidate 
model structures 
because unlike the score (which is an average of $E_\text{val}$ values), 
$E_\text{test}$ is not biased towards fits that work especially well for the 
validation samples but not necessarily to other samples. It can therefore
serve as an indicator of the significance of the difference in scores between
candidate model structures.

\begin{figure}
    \includegraphics[width=0.49\textwidth]{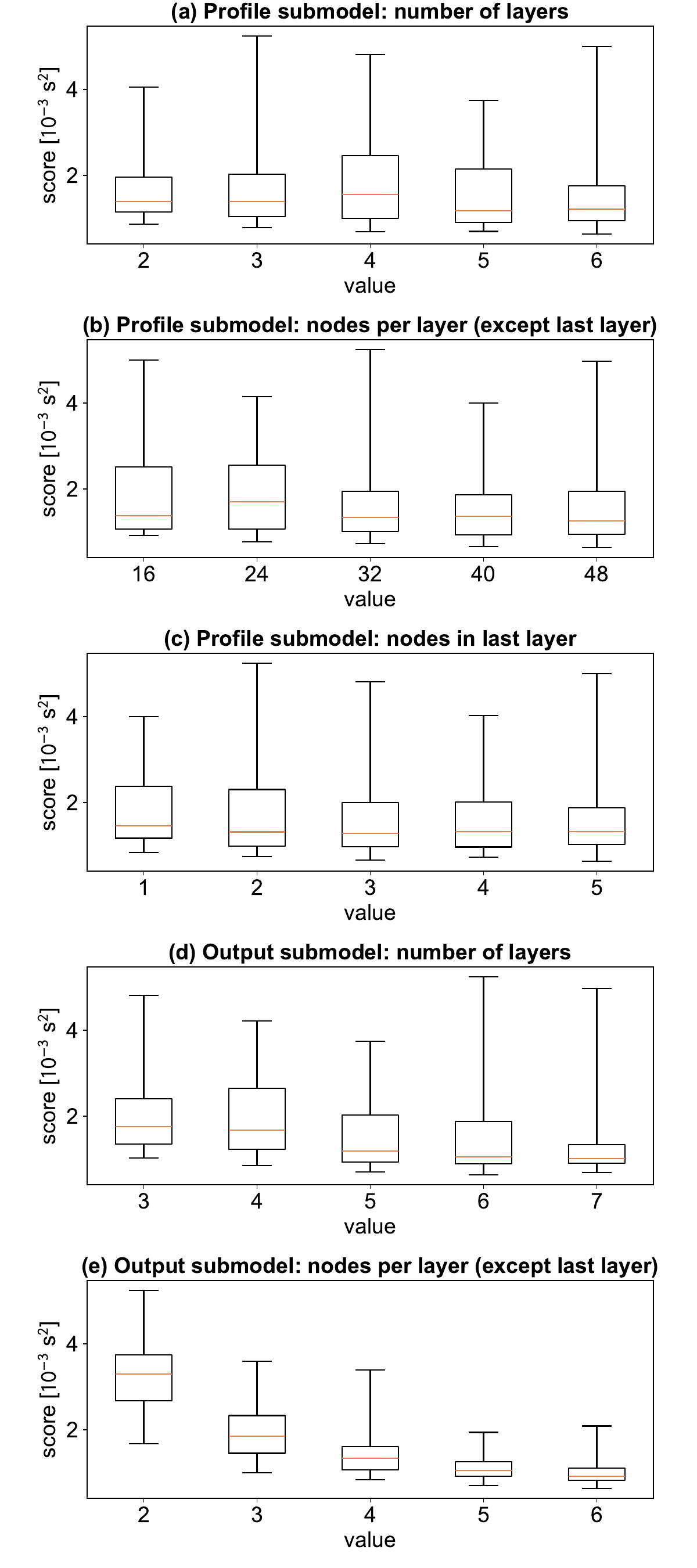}
    \caption{Box plots illustrating distributions of the scores achieved by
             a set of candidate models, grouped by the values of the 
             hyperparameter indicated in the title of the corresponding panel.
             The line in the middle of each box represents the median value;
             the bottom and top of the box represent the twenty-fifth and
             seventy-fifth percentile, respectively; and the whiskers extend
             to the minimum and maximum values.}
    \label{fig:tuner_stats}
\end{figure}

Fig.~\ref{fig:tuner_stats} shows an overview of the relative performance of
model structures with specific values of each of the hyperparameters. 
As an example, in Fig.~\ref{fig:tuner_stats}a, the 320 candidate structures 
under consideration
are grouped according to the number of layers in the profile submodel,
delineated on the x-axis. For each group, a vertically-oriented box plot is 
displayed to show the distribution of scores obtained for candidate 
models within the group. The bottom and top of each box correspond to the
twenty-fifth and seventy-fifth percentiles of the scores, respectively;
the line in the middle of the box represents the median; and the wiskers
below and above the box extend to the minimum and maximum values.
Figs.~\ref{fig:tuner_stats}b-e show distributions of the scores for the
candidate model structures regrouped according to values of each of the other
hyperparameters.

From the comparisons in Fig.~\ref{fig:tuner_stats}, it appears that the 
model suitability was most sensitive to the number of nodes per layer in the 
output submodel (Fig.~\ref{fig:tuner_stats}e). The median and quartiles of
the score all decreased significantly with each increment of this
hyperparameter value. To a lesser degree, increases in the number of layers
in the output submodel (Fig.~\ref{fig:tuner_stats}d)
are also associated with reductions in the score. By contrast,
the discrepancies in scores between candidate structures with different values
of hyperparameters related to the profile submodel 
(Fig.~\ref{fig:tuner_stats}a-c) are less significant.
Notably, the model suitability apparently depends little on 
the number of features extracted from the profiles to feed into the output
submodel (\textit{i.e.} the number of nodes in the last layer, 
Fig.~\ref{fig:tuner_stats}c). According to the distributions, many models
that only extracted one feature from the profiles achieved results that were
just as accurate as models that extracted five features.

Overall, the search revealed that candidate model structures with many 
different 
combinations of hyperparameters could produce predictions with similarly low 
scores. The top twenty models sampled had scores ranging from 
$6.43\times10^{-4}$~s${^2}$ to $8.27\times10^{-4}$~s${^2}$.
The $E_\text{test}$ value for the same candidate structures ranged from 
$7.15\times10^{-4}$~s${^2}$ to $11.8\times10^{-3}$~s${^2}$ with little 
correlation with the corresponding scores. Hence, the top twenty candidate 
structures do not differ significantly in their prediction accuracy.
From these top twenty, we selected the lowest-scoring structure with 
fewer than 3,000 trainable parameters (weights), favoring its simplicity 
relative to the first-place structure, which had more than 10,000 weights. 
This is the structure depicted in Fig.~\ref{fig:nn}.
We will use this model structure for the analysis in the rest of the paper.
However, based on the analysis from the tuner search, it should be noted that 
many different model structures could likely yield similar results.

\subsection{Model evaluation}
\label{sec:model_evaluation}

To evaluate the attainable prediction accuracy in greater depth, we fit
the model with the structure outlined in Fig.~\ref{fig:nn} separately to
sample sets from the high-iota configuration (which was also used for the
structure evaluation in Sec.~\ref{sec:model_structure}) and the standard 
configuration.
For these fits, each sample set was randomly partitioned into two subsets:
a training set used for training the model (80\% of the samples) and a 
test set to evaluate model performance (20\% of the samples).

\begin{figure*}
    \includegraphics[width=\textwidth]{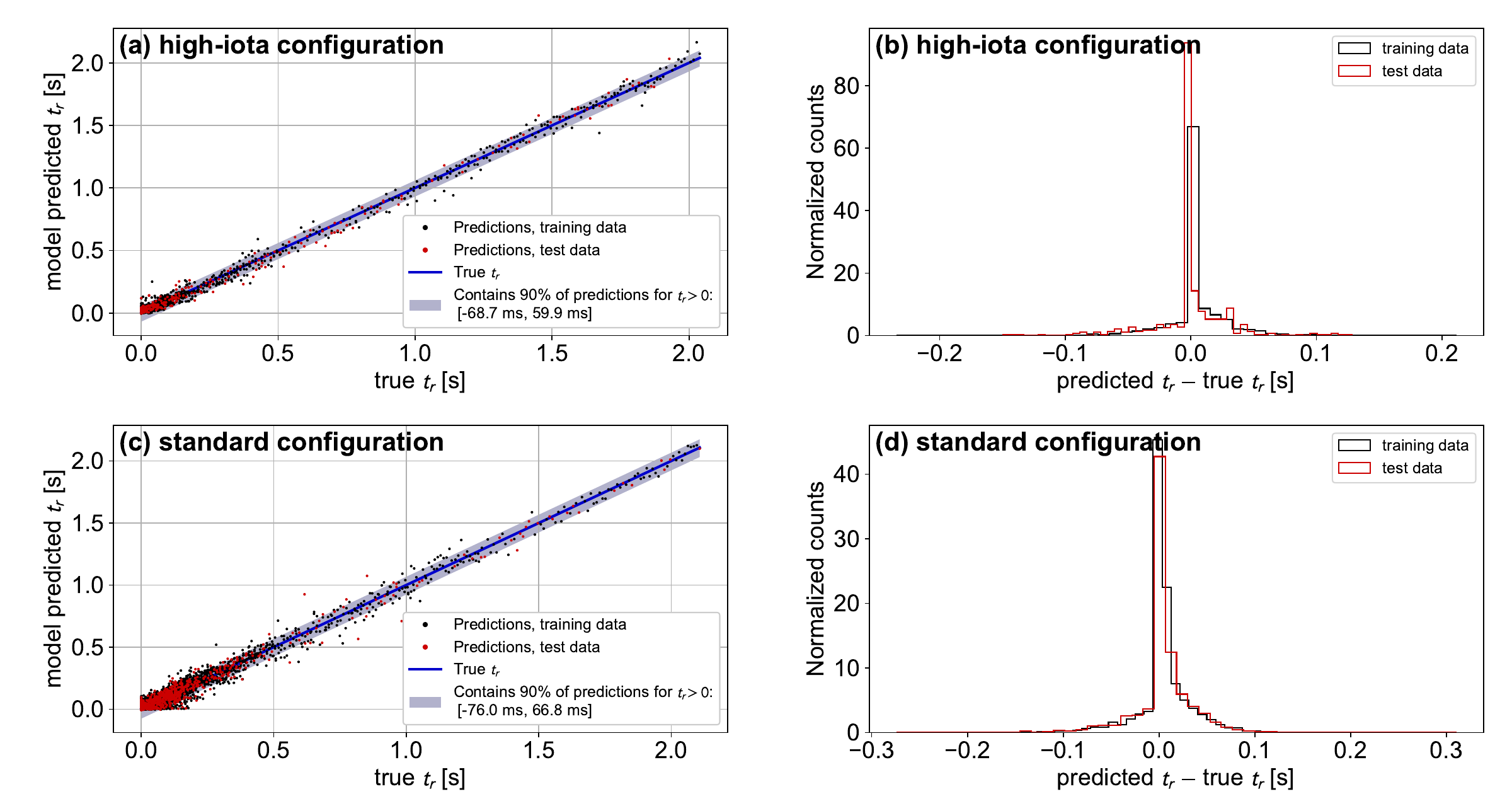}
    \caption{Quantification of (a) prediction accuracy and (b) error 
             distribution for models trained on data samples from the 
             high-iota configuration, and likewise for the standard 
             configuration in (c) and (d).
             (a) and (c): Comparison of predictions of $t_r$ with true values 
                of $t_r$. Black dots indicate predictions for data samples
                used to train the model; red dots indicate predictions for
                samples in the test set that were not used for training. 
                The light blue stripes enclose 90\% of the predictions for 
                samples with true $t_r$ values greater than 0.
             (b) and (d): Normalized histograms showing the distributions of 
                discrepancies between the predicted $t_r$ and true $t_r$ for
                the training data (black) and the test data (red).}
    \label{fig:model_assessment}
\end{figure*}

The prediction accuracy of
the models is shown in Fig.~\ref{fig:model_assessment}. 
In Fig.~\ref{fig:model_assessment}a and \ref{fig:model_assessment}c, model 
predictions for $t_r$
are plotted against the true values for the high-iota and standard configuration
sample sets, respectively.
Black data points indicate samples used to train the models, whereas red points
indicate samples from the test sets that were excluded from training.
Perfect predictions would fall on the solid blue line on each plot.

Fig.~\ref{fig:model_assessment}b and \ref{fig:model_assessment}d show the 
distributions of the errors
in predictions for the training and test samples in the high-iota configuration
and the standard configuration, respectively. The distributions are represented
as normalized histograms, with the training samples illustrated in black and
the test samples illustrated in red. For both configurations, the distributions
of training and test data are quite similar, hence there is no indication that
the models were over-fit to the training data.

The accuracy of the predictions varied depending on whether the samples
were taken during the phase of enhanced confinement (when the true value
of $t_r$ decreases linearly with time; see Fig.~\ref{fig:tr_definition}b) 
or after enhanced confiment is lost (when $t_r=0$). During the 
ehanced-confinement phase, at least 90\% of the model-predicted values of 
$t_r$ fell within 75 ms of the true values. This range was similar for
both models and is illustrated for each by the light-blue stripes in
Fig.~\ref{fig:model_assessment}a,c. During the phase after the loss 
of enhanced confinement, the accuracy was substantially better. Both
models predicted values of $t_r$ within 30~ms of the correct value of 0 ms
for at least 90\% of the samples from this phase. Taking both phases together, 
the fifth and
ninety-fifth percentiles of the prediction errors were $(-39$ ms$,~43$ ms$)$ 
for the high-iota configuration samples and $(-44$ ms$,~51$ ms$)$ for the 
standard configuration samples.

This level of accuracy is more than sufficient to be useful for control 
purposes, as the typical error in all cases is substantially less than the 
energy confinement times typically observed 
in W7-X ($> 100$ ms) as well as the minimum temporal spacing between 
subsequent pellet injections ($\sim 100$ ms). 
Furthermore, the model can make predictions of $t_r$ quickly enough to be useful
for real-time applications. The average time to evaluate the model
for a single sample is less than 5~$\mu$s on a single CPU. By comparison, 
profiles of $T_e$ and $n_e$ are only acquired roughly every 11 ms by the 
Thomson scattering system in normal operation.

\begin{figure}
    \includegraphics[width=0.49\textwidth]{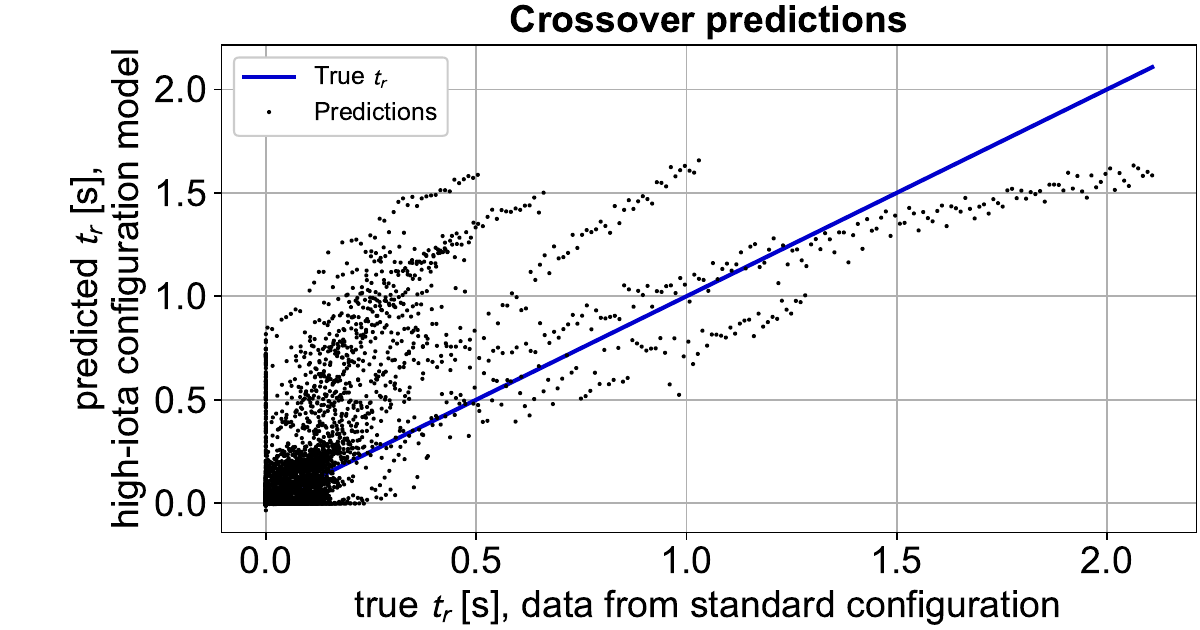}
    \caption{Predictions of $t_r$ made using the model trained for the high-iota
             configuration applied to data obtained from the standard 
             configuration.}
    \label{fig:crossover_predictions}
\end{figure}

While the models exhibit high accuracy for data samples from their respective
magnetic configurations, the two models are substantially different from one
another. This is illustrated in Fig.~\ref{fig:crossover_predictions}, which
shows predictions of $t_r$ made by the high-iota configuration model for data 
samples from the standard configuration, plotted against the true $t_r$ values.
Here, the errors are much higher than what is attained when each model makes
predictions for its intended magnetic configuration. This indicates the 
impact of the magnetic configuration on the relationship between input
data and $t_r$.

\section{Potential for use in feedback applications}
\label{sec:feedback_potential}

To further evaluate the model performance and to illustrate its potential 
utility for future experiments, we will now apply the trained models to two 
examples of discharges in which plasma performance was enhanced with series
of pellets. We note that the discharges shown in this section were not included 
in the model training data. 

\begin{figure*}
    \includegraphics[width=\textwidth]{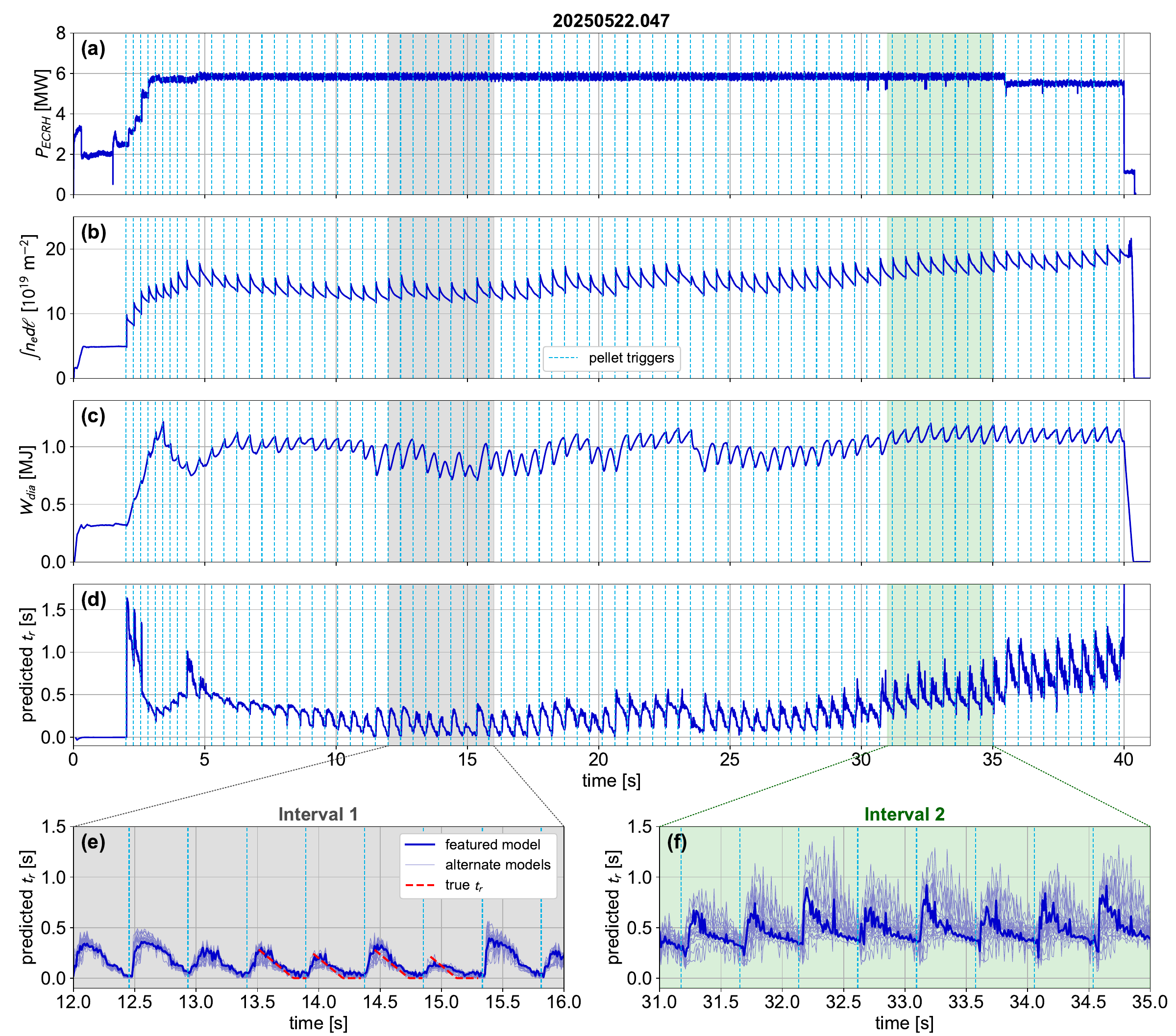}
    \caption{Key time traces from a plasma discharge for which $t_r$ was 
             computed based on recorded experimental data:
             (a) ECRH power,
             (b) line-integrated electron density,
             (c) diamagnetic energy,
             (d) $t_r$ predicted by the model,
             (e) expanded view of the $t_r$ prediction for the gray-shaded 
                 interval (Interval 1) along with predictions by alternate
                 models, and
             (f) expanded view of the $t_r$ prediction for the green-shaded 
                 interval (Interval 2) along with predictions by alternate
                 models.
            }
    \label{fig:use_case_low_rate}
\end{figure*}

The first example discharge is shown in Fig.~\ref{fig:use_case_low_rate}.
This discharge was heated purely by ECRH and featured a long series of pellet
injections lasting approximately 38 seconds. As with all pellet injections
to date in W7-X, the injection times were pre-programmed. From time 
$t=2$~s through $t=4.3$~s, pellets were injected approximately every 0.28~s, 
or at a rate of 3.6~Hz, as the heating power was stepped up to a final value
of 6 MW. For the rest of the discharge, the pellet injection rate slowed down
to once every 0.48~s (2.1~Hz), while the heating power remained constant.

While the key control parameters (heating power level and pellet injection rate)
remained constant from 5~s to 35~s, the confinement quality varied
over time as seen in the trace of $W_{dia}$ in 
Fig.~\ref{fig:use_case_low_rate}c. Overall, $W_{dia}$ fluctuates throughout
the discharge in response to pellet injections. However, there are some 
notable trends in $W_{dia}$ that average out over intervals of several 
pellets. For example, from $t=12$~s to $t=16$~s (Interval 1, shaded in gray in 
the plots), 
$W_{dia}$ has a time-averaged value of 0.85~MJ. By contrast, from $t=31$~s to 
$t=35$~s (Interval 2, shaded in green), it has a time-averaged value of 
1.10~MJ, about 25\% higher. 

Model predictions for $t_r$ for this discharge are shown in 
Fig.~\ref{fig:use_case_low_rate}d, with expanded views in 
Fig.~\ref{fig:use_case_low_rate}e-f for the data in the gray- and 
green-shaded intervals. These values were determined by evaluating the
trained model for profiles of $T_e$ and $n_e$ calculated from recorded
data from the Thomson scattering diagnostic, as well as interpolated values of
the total ECRH power at the corresponding time points. As this discharge was 
performed in the high-iota magnetic configuration with O2-mode ECRH, 
predictions of $t_r$ were made with the model trained on plasma data from
these conditions.

Within Interval 1, there were four time intervals during which $\tau_E$ clearly
settled to a flat level indicating a relaxation to a lower confinement 
state. For these intervals, it was thus possible to calculate a true value of 
$t_r$. The true $t_r$ values for these intervals are indicated as the 
red dashed lines in Fig.~\ref{fig:use_case_low_rate}e.

For an indicator of the robustness of the model predictions, we trained 19
additional models on the data from the high-iota configuration. Each model
used the same structure as the ``featured model'' whose predictions are plotted
in dark blue in Fig.~\ref{fig:use_case_low_rate}d-f,
but the optimizer was trained on a different randomly-selected subset of 80\% 
of the samples from the high-iota sample set. The non-deterministic nature of 
the optimizer
used for fitting also contributed to the variability in the models. Predictions
by these alternate models are shown as pale blue time traces in 
Fig.~\ref{fig:use_case_low_rate}e-f. 

During Interval 1, the alternate models exhibit relatively little variation 
from the predictions of the featured model. 
By contrast, in Interval 2, there is much more variation in the predictions,
with minimum and maximum predictions separated by 750~ms or more in some cases.
It is postulated that the reason for this is that the training data did not
include samples from plasma scenarios like those depicted in Interval 2,
with high ECRH heating power and enhanced performance maintained steadily
with pellets. By construction, it was not possible to obtain samples from
such cases because the plasma never exited the enhanced confinement state
and as a result $t_\text{end}$ was not known for the purpose of determining true
values of $t_r$.

Overall, the qualitative behavior of the $t_r$ time trace indicates that the
model is working as expected. Prior to the onset of pellets
(for $t<2$~s), the model estimates $t_r$ of zero or nearly zero, which is
the correct value for a plasma that is not exhibiting enhanced 
performance. 
Following the onset of pellet injections, $t_r$ takes on a
sawtooth-like form, rapidly rising after each pellet injection and then 
steadily decreasing until the next pellet injection. This is roughly
the expected behavior of the model. In the intervals between pellet injections,
when the plasma is in a state of enhanced confinement with constant heating
power and no other major external interventions, the model-predicted $t_r$
should ideally decrease linearly with time as shown in 
Fig.~\ref{fig:tr_definition}b. Then, each time a pellet is injected,
it refuels the core of the plasma and restores the peaked density profile
that had been gradually decaying since the previous pellet was injected.
Hence, it is to be expected that $t_r$ should jump rapidly, as the latest
pellet has quickly adjusted the plasma profiles in a way that extends the
phase of enhanced confinement. 
We also note that the agreement of 
the predictions of all models is quite good with the true $t_r$ values where
they are known (Fig.~\ref{fig:use_case_low_rate}e).

Upon closer inspection, it is clear that the model does not behave exactly
according to its ideal definitions. This can be seen in the expanded views of 
the $t_r$ traces in Fig.~\ref{fig:use_case_low_rate}e-f.
For example, there is a finite time delay on the order of $50$~ms between the 
injection of a pellet (approximated by the trigger times represented by the
vertical cyan dashed lines) and the resulting rise in $t_r$. Furthermore, 
once the prediction for $t_r$ reaches its post-pellet peak value, the 
subsequent decrease in time is not exactly linear with a slope of $-1$ as 
stipulated in the model definition (Eq.~\ref{eqn:t_r}), although it is often
close. 

In spite of these non-ideal aspects, the model predictions of $t_r$ exhibit
physically meaningful features that could be useful for a future control system.
First, there are important quantitative differences between the predicted
$t_r$ for Interval 1, which had a lower time-averaged $W_{dia}$ 
(Fig.~\ref{fig:use_case_low_rate}e, gray stripe) and Interval 2, which had a
higher time-averaged $W_{dia}$ (Fig.~\ref{fig:use_case_low_rate}f, green 
stripe). During Interval 1, $t_r$ rarely exceeds 0.48~s, which is the temporal
spacing between pellets. Accordingly, it often drops to zero before the
subsequent pellet is injected, indicating that the plasma has lost the 
enhancement in confinement brought about by the previous pellet (see the
definitions in Eq.~\ref{eqn:t_r} and Fig.~\ref{fig:tr_definition}).
By contrast, during Interval 2, although there is substantial variability 
between
the predicted $t_r$ from different models, all models predict $t_r$ to exceed 
0.4~s nearly the entire time
and to never reach 0. This is consistent with the higher time-averaged 
$W_{dia}$ of this interval and indicates that the plasma remains in a state
of enhanced confinement the entire time with no temporary dropoffs.

The differences between the behavior of $t_r$ in Intervals 1 and 2, and the
corresponding differences in stored energy, indicate how the $t_r$ model
can be useful as a predictive tool for a control system. During Interval 1
(Fig.~\ref{fig:use_case_low_rate}e), the values of $t_r$ at any given time
are mostly less than the amount of time before the next pellet is scheduled to 
arrive. This indicates that the model is predicting that the
plasma will lose its enhanced confinement before the next pellet. 
If the pellets had been controlled by a feedback system with access to
these $t_r$ predictions in real time, the feedback system could have responded
by increasing the pellet injection rate to avoid losing enhanced confinement.
For example, if the predicted value of $t_r$ was 0.3~s, the control system 
would know that it should inject a pellet within the next 0.3~s to preserve
the state of enhanced confinement.

During Interval 2 (Fig.~\ref{fig:use_case_low_rate}f), by contrast, $t_r$ is
consistently greater than the amount of time before the next scheduled pellet.
Hence, a feedback control system equipped with real-time $t_r$ estimates 
would have the knowledge that it would not be necessary to 
increase the pellet injection rate to maintain enhanced confinement. In 
fact, the predictions of $t_r$ are high enough that the the control system 
could afford to reduce the injection rate while still maintaining enhanced
confinement if there are other motivations for doing so.

Overall, the results in Fig.~\ref{fig:use_case_low_rate} give us reason to
believe that real-time estimates of $t_r$ have the potential to improve overall 
performance if used as an input to a feedback
system controlling the pellet injector.
In particular, estimates of $t_r$ produced during phases of 
enhanced confinement appear to give reasonably accurate predictions of 
how much time remains before the enhanced confinement is lost. These predictions
can be made with sufficient advance notice for the control system to react 
by injecting a pellet before the loss of enhanced confinement is predicted to
occur, thereby keeping the plasma consistently in a state of enhanced 
confinement.

\begin{figure}
    \includegraphics[width=0.49\textwidth]{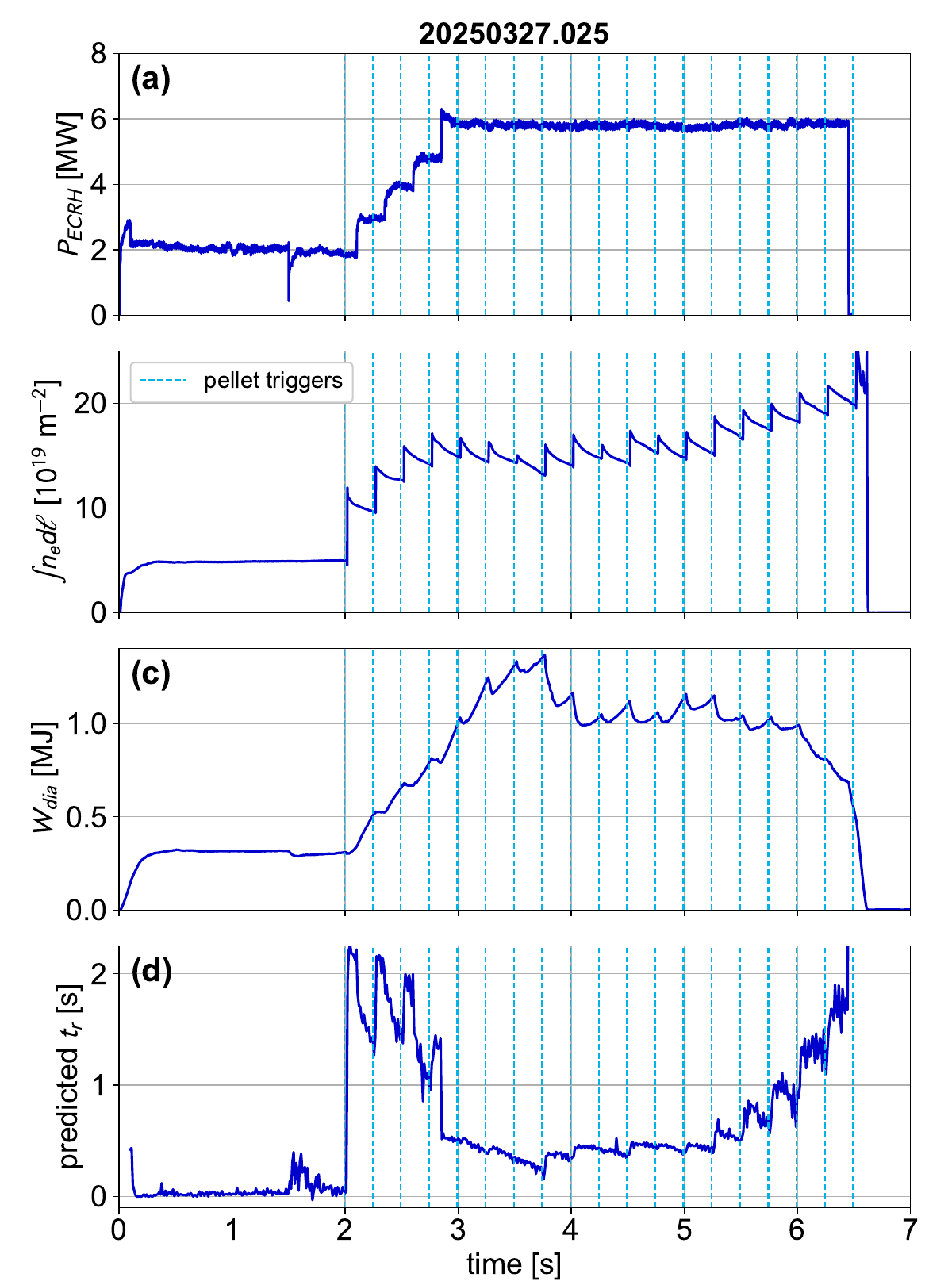}
    \caption{Key time traces from a plasma discharge for which $t_r$ was 
             computed based on recorded experimental data:
             (a) ECRH power,
             (b) line-integrated electron density,
             (c) diamagnetic energy, and
             (d) $t_r$ predicted by the model.
            }
    \label{fig:use_case_high_rate}
\end{figure}

In addition to helping to optimize plasma performance, real-time estimates of
$t_r$ could help to avoid overfueling and premature discharge termination.
An example of such a scenario is shown in Fig.~\ref{fig:use_case_high_rate}.
Here, pellets were injected at a constant, pre-programmed rate of 4~Hz (one 
pellet every 0.25~s), beginning at $t=2$~s when the heating power was stepped
up. During the portion of the discharge with 6~MW of heating power and 
consistent pellet injection (Fig.~\ref{fig:use_case_high_rate}a), overall 
energy confinement remained high for
several seconds, with $W_{dia}$ exceeding 1 MJ 
(Fig.~\ref{fig:use_case_high_rate}c)---substantially greater than
what would be achievable in absence of pellet injection. However, around
$t=5$~s, the plasma density begins to increase steadily 
(Fig.~\ref{fig:use_case_high_rate}d) until the discharge is terminated
due to increasing levels of stray ECRH radiation. The heightened level of stray 
radiation likely results from excessively high density: either the 
corresponding drop in electron temperature led to poor absorption of the O2-mode
or the density rose above the cutoff level for O2-mode propagation. 

Model predictions of $t_r$ shown in Fig.~\ref{fig:use_case_high_rate}d show 
a potential strategy for avoiding such an unplanned termination in the future
while maintaining good performance. Earlier in the discharge (particularly
around $t=3.75$~s), the predicted $t_r$ falls as low as 0.2~s, indicating that
the relatively fast pellet injection rate of 4~Hz was likely necessary to 
avoid losing good confinement. However, 
after $t=5$~s, the predicted $t_r$ rises past 0.5~s, indicating
the pellet injection rate could have slowed down substantially, to 2~Hz or 
lower, without a loss of enhanced confinement. Lowering the injection rate 
would have resulted in a lower overall fueling rate, which would have reduced
the likelihood of the uncontrolled rise in density that occurred with the
4~Hz injection rate. Hence, a feedback control system with the ability
to adjust the injection rate based 
real-time $t_r$ predictions could improve both the reliability and 
performance of pellet-fueled
discharges.

\section{Discussion}
\label{sec:discussion}

Based on the promising results obtained from the neural network models 
discussed in this paper, we anticipate that the 
remaining-time predictor can be useful for plasma profile control experiments,
either as a sole input or as part of a more sophisticated model-predictive
control system. One simple application would be to program a feedback 
controller to inject a pellet any time $t_r$ falls below a certain 
user-specified minimum value. In principle, this would ensure that the state
of enhanced confinement is never lost, leading to better time-averaged
energy confinement times and stored energy than what has been achieved to
date in W7-X with pre-programmed pellet injection times. Also, such a control
system may naturally avoid raising the plasma density above the cutoff
level, as such incidents are typically preceded by sustained increases
of $t_r$ to high levels that would be well above a reasonable minimum
(as shown in Fig.~\ref{fig:use_case_high_rate}).

The model could also serve as an input to more sophisticated plasma profile
control algorithms that seek to optimize performance. 
A model-predictive pellet control 
algorithm, along the lines of algorithms currently under development for ITER 
and other tokamaks \cite{bosman2023a,orrico2025a}, could use $t_r$ estimates
to impose a constraint on its choice for the pellet injection rate as it
seeks to meet more sophisticated targets, such as specific profile gradients.

While the models studied in this paper already exhibit sufficient accuracy
to be useful
in at least some experimental contexts, there are many opportunities to expand
the model to improve accuracy. For one, the model could be expanded to take
inputs from additional diagnostics that may provide information not encoded
in the temperature and density profiles or in the heating power level.
Real-time data from neutral pressure gauges \cite{wenzel2022a}, for example, 
would allow the 
model to explicitly account for the edge neutral pressure and its possible
impacts on core transport. In addition, the X-ray Ion Crystal Spectrometry
(XICS) diagnostic \cite{langenberg2018a,pablant2018a} is currently being 
upgraded to produce ion temperature
profiles in real time, which may also be informative for model predictions.

One notable finding from the hyperparameter search described in 
Sec.~\ref{sec:model_structure} was that the predictive capability of the
model was not sensitive to the number of nodes in the last layer of the 
profile submodel. 
Model structures with a single node in this layer could achieve prediction 
accuracy similar to that of model structures with five, as shown in
Fig.~\ref{fig:tuner_stats}c. This suggests that the information encoded in
the combined electron temperature and density profiles that is relevant
for predicting $t_r$ can be reduced to a single scalar feature.
The nature of this feature is not understood but will be the subject of 
future analysis.

\section{Conclusion}
\label{sec:conclusion}

In summary, we have presented the development and testing of a data-driven
model that makes rapid predictions of impending loss of plasma confinement
according to real-time available diagnostic data. The key inputs to the model
are profiles of electron temperature and electron density from eight spatial
locations observed by the Thomson scattering diagnostic, as well as the total
ECRH heating power. The output of the model is an prediction of the remaining
time, $t_r$, before the plasma will relax out of a state of enhanced 
confinement without an external intervention such as the injection of a pellet.

The model is implemented as an artificial neural network trained on past 
experimental data. To accommodate the different magnetic configurations 
employed in W7-X experiments, separate, dedicated models are trained on 
data from each respective configuration. Models trained based on data for the
standard and high-iota magnetic configurations each were able predict
$t_r$ during periods of enhanced confinement to within 75~ms or better for
90\% of the data samples considered. The models were also able to recognize
when the plasma was \textit{not} in a state of enhanced confinement with 
even better accuracy; specifically, for samples with a true $t_r$ value of 0~s,
the models were both within 30~ms of the correct value for at least 90\% of
the tested samples. Considering all samples together, both models were within
51~ms of the correct value for at least 90\% of the samples.
This level of discernment is expected to be adequate
for control purposes, as the energy confinement time and the temporal spacing
between pellets are both substantially greater than 75~ms on W7-X. Furthermore,
the model can be evaluated quickly for use in real-time applications: each 
evaluation of $t_r$ takes less than 5 $\mu$s on a CPU.

Finally, we used the trained models to compute $t_r$ for exemplary plasma
discharges to demonstrate what $t_r$ would look like if implemented as a 
real-time signal. Traces of $t_r$ indicate that it could provide useful
information for a plasma control system, including timely predictions of
when a pellet should be injected to avoid a loss of confinement enhancement.
Such predictions could be used adjust the pellet injection rate throughout
a discharge to strike a balance between maintaining enhanced confinement and
avoiding overfueling.

The remaining-time predictor developed in this work is among a growing number
of neural network-based models that can make rapid and accurate predictions
of plasma behavior in mangetic fusion experiments 
\cite{abbate2021a, seo2024a, jalalvand2025a}. It does not contain any
specific structure or constraints based on physical models, but rather
is based solely on empirical observations from previous experiments.
Hence, the models employed in this work that were trained on W7-X data
may not be directly applicable to other magnetic fusion devices. On the
other hand, it seems reasonable to expect that similar models could be 
trained for other devices using their respective experimental data and 
utilized to improve performance and reliability in similar ways to what is
anticipated for the W7-X models.

The fact that models can be trained to make reasonable predictions for the
duration of enhanced confinement based primarily on profile measurements
is an indicator that the evolution of plasma
profiles following pellet injections follows consistent physical processes.
Efforts are currently underway to employ 3D neoclassical, gyrokinetic, and
transport codes to compute local heat fluxes and the resulting plasma profile
evolution from first principles. In the meantime, empirical, data-driven
models like the one developed in this work can play an important role in
bridging the gap between physics modeling capabilities and the needs for
rapid predictions for plasma profile control.

\section*{Acknowledgments}

This work was supported by the U.S. Department of Energy under Contract 
No.~DE-AC02-09CH11466 and Grant No.~DE-SC0014229.
The United States Government retains a non-exclusive,
paid-up, irrevocable, world-wide license to publish or reproduce the published 
form of this manuscript, or allow others to do so, for United States 
Government purposes.
The work was carried out within the framework
of the EUROfusion Consortium, funded by the European Union via the Euratom
Research and Training Programme (Grant Agreement No.~101052200 -- EUROfusion).
Views and opinions expressed are those of the authors only and do not
necessarily reflect those of the European Union or the European Commission.

\end{document}